\documentclass[aps,twocolumn,pra]{revtex4-1}
\usepackage{newlfont}
\usepackage{color}           % color to the text 
\usepackage{soul}            % strickthrough the text
\usepackage{amssymb}
\usepackage{amsfonts}
\usepackage{amsmath}
\usepackage{wasysym}
\usepackage{graphicx}
\usepackage{epsfig}
\usepackage{amsthm}
\usepackage{bm}
\usepackage[colorlinks=true, citecolor=blue, urlcolor=blue ]{hyperref}

\begin{document} 
  
\title{
% Heisenberg limited metrology with separable Gaussian states and a new measure of quantum correlation} 
Quantum Fisher information as the measure of Gaussian quantum correlation: \\ Role in quantum metrology} 
% \title{Quantum Dynamical Quantification of Non-Classical Correlation}
%{Quantum Correlation Dictates the Speed of Quantum Evolution?}
\author{Manabendra N Bera}
\affiliation{Harish-Chandra Research Institute, Chhatnag Road, Jhunsi, Allahabad 211 019, India}

\begin{abstract} 
We have introduced a measure of Gaussian quantum correlations based on quantum Fisher information. For bipartite Gaussian states the minimum quantum Fisher information due to local unitary evolution on one of the parties reliably quantifies quantum correlation. In quantum metrology the proposed measure becomes the tool to investigate the role of quantum correlation in setting metrological precision.
% how quantum correlations are influential in modulating metrological precision. 
In particular, a deeper insights can be gained on how quantum correlations are instrumental to enhance metrological precision. 
% the instrumental role of quantum correlation to enhance metrological precision. 
Our analysis demonstrates that not only entanglement but also quantum correlation plays an important role to enhance metrological precision. Clearly unraveling the underlaying mechanism we show that quantum correlations, even in the absence of entanglement, can be exploited as the resource to beat standard quantum limit and attain Heisenberg limit in quantum metrology. 
% 
% 
% gain deeper insights on how quantum correlations are influential in modulating metrological precision. In particular 
% 
% 
% the role of quantum correlation that modulates the metrological precision. 
% 
% 
% To investigate the role of quantum  correlation in quantum  metrology, the proposed measure establishes itself as the 
\end{abstract}

\maketitle

% \textit{Introduction}  -- 
Quantum correlations are the most esoteric property in quantum world. They have been subjected to intensive studies, in the past two decades, mainly from the belief that they are fundamental resource for quantum technologies \cite{Braunstein05, Horodecki09}. 
% information precessing tasks and quantum metrology. 
Initially, entanglement was considered to be the only form of quantum correlation and was thought to play central role in various applications e.g., in achieving better precision in quantum metrology \cite{Paris09} and in quantum communication tasks \cite{Horodecki09}. However, several observations such as the requirement of very little entanglement for certain quantum information and computation tasks \cite{Datta05, Nest12} indicate that there may present quantum correlations beyond entanglement, which are in action. A more general quantum correlation measure \cite{Vedral12} --quantum discord \cite{Ollivier01, Giorda10}-- was introduced for, both discrete and continuous variable, bipartite systems  and  it has been proposed as the resource for certain quantum computation tasks \cite{Datta08}, encoding information onto a quantum state \cite{Gu12}, and quantum state merging \cite{Madhok11}.
% Researchers confirm that there do exist quantum correlations, with no classical analogue, in separable (i.e., not entangled) states that has potential applications in quantum technologies \cite{Datta08}. 
In the context of quantum metrology with discrete variable systems, it has been shown that the presence of quantum correlation guarantees a non-vanishing metrological precision \cite{Girolami14}. Recently, the role of quantum correlation is studied in the context of metrology \cite{BeraQFICorr}. Clearly unearthing the mechanism behind, it has been shown that quantum correlation can be exploited to enhance precision beyond standard quantum limit (SQL) and to attain Heisenberg limit (HL) even in the absence of entanglement \cite{BeraQFICorr}.
% 
% 
% it is revealed that quantum correlation can increase interferometric power \cite{Girolami14}. Further it has been shown that, even in the absence of entanglement, quantum correlation can enhance precision beyond standard quantum limit (SQL) even to attain Heisenberg limit (HL) \cite{BeraQFICorr}.
% Recent studies on metrology with discrete variable system reveal that, in the absence of entanglement, quantum correlation can enhance precision beyond standard quantum limit (SQL) even to attain Heisenberg limit (HL). 
For metrology with continuous variable (CV) systems, several observations indicate the affirmative role of quantum correlation to enhance metrological precision (see for example \cite{Sahota14}). However a detailed understanding on the underlaying mechanism is still lacking. 

In this paper we endeavor to fill this gap. To attack the problem, we introduce a measure of Gaussian quantum correlation based on quantum Fisher information (QFI). For bipartite Gaussian states the minimal QFI over all local unitary evolution quantifies the quantum correlation, reliably. The proposed measure can  be interpreted, from quantum dynamical perspective, as the minimum quantum speed of evolution due to local unitaries. This quantum dynamical aspect of quantum correlation provides us the premise to investigate how quantum correlation is instrumental in quantum metrology. Our analysis shows that even for separable Gaussian states (i.e., not entangled) the quantum correlation can play decisive role to reduce metrological error and attain HL. 
% A tighter bound on metrological error is also derive both in the presence and absence of quantum correlation. 

% \textit{ Quantum Fisher information (QFI)} -- 
Our discussion is centralized on quantum Fisher information (QFI) which is an important quantity in quantum geometry of state spaces \cite{Wootters81, Petz96}, quantum information theory \cite{Bengtsson06} and metrology \cite{Paris09}. In particular, it quantifies the quantum speed of a smooth evolution where the distance in the quantum state space is measured in terms of Bures distance and the inverse of QFI delimits the precision in quantum metrology. For the given quantum states $\rho(t_1)$ and $\rho(t_2)$, parameterized by time $t_1$ and $t_2$, the Bures distance is defined as $S_{Bu}^2=4\left(1-\mbox{Tr}\sqrt{\sqrt{\rho(t_1)}\rho(t_2)\sqrt{\rho(t_1)}}  \right)$. If these two quantum states are connected through a smooth dynamical process, say unitary evolution $U=\mbox{exp}(-iHt)$, the geometric distance $dS_{Bu}^2$ due to infinitesimal time translation from $t$ to $t+dt$ is given by
\begin{equation}
 dS_{Bu}^2=\mathcal{F}^2(\rho(t),H)dt^2
\end{equation}
up to the second order in $dt$. The quantity $\mathcal{F}(\rho(t),H)$ is QFI and signifies the quantum speed of evolution $\frac{dS_{Bu}}{dt}$. In terms of symmetric logarithmic derivative (LSD) operator, $L(t)$,  QFI is expressed as $\mathcal{F}^2(\rho(t),H)=\mbox{Tr}[\rho(t)L(t)^2]$. The $L(t)$ is defined implicitly by $\frac{d\rho(t)}{dt}=\dot{\rho}=\frac{1}{2}\{L(t),\rho(t)\}$ where the curly braces denote anticommutator. Thus  QFI becomes \cite{Braunstein94}
% $\mathcal{F}^2(\rho(t),H)=\sum_{ij}\frac{|\dot{\rho}_{ij}|^2}{\rho_{ii}+\rho_{jj}}$.
\begin{equation}
 \mathcal{F}^2(\rho(t))=2\sum_{ij}\frac{|\dot{\rho}_{ij}|^2}{\rho_{ii}+\rho_{jj}},
\end{equation}
where the summands are taken with the condition $(\rho_{ii}+\rho_{jj})>0$.
In general, deriving QFI for continuous variable system is quite cumbersome. Very recently researchers have made the breakthrough to derive its analytical form for Gaussian systems \cite{Jiang14}. 

% Before we proceed further to outline the derivation of QFI, we briefly discuss about Gaussian states. 
An $N$-mode Gaussian system is characterized by $2N$ canonical degrees of freedom. Let $x_k$ and $p_k$ be the ``position'' and ``momentum'' operators of the $k$-th mode with $k=1,...,N$, acting on the associated Hilbert space $\mathcal{H}_N$. Defining $r=(x_1,..,x_N,p_1,...,p_N)^T$, the canonical commutation relation can be written as 
\begin{equation*}
 [r_k,r_l]=iJ_{kl} \ \   \mbox{where} \ \ 
 J=\oplus_{1}^n\begin{pmatrix}
  0 & \mathbb{I} \\
  -\mathbb{I} & 0
   \end{pmatrix} =-J^T=-J^{-1}
\end{equation*}
with $\mathbb{I}$ being the $N \times N$ identity matrix.
% Defining $r=(x_1,p_1,...,x_N,p_N)$, the canonical commutation relation can be written as 
% \begin{equation*}
%  [r_k,r_l]=iJ_{kl} \ \   \mbox{where} \ \ 
%  J=\oplus_{1}^n\begin{pmatrix}
%   0 & 1 \\
%   -1 & 0
%    \end{pmatrix}
% \end{equation*}
A Gaussian state $\rho$ is uniquely characterized by its first moment $d=(d_1,...,d_{2N})\in\mathbb{R}^{2N}$ with $d_k=\mbox{Tr}(\rho r_k)$ and its second moment, the covariance matrix, $[\Gamma]_{kl}=\mbox{Tr}(\rho\{ r_k-d_k,r_l-d_l  \})$, where the curly braces denote anticommutator. In terms of $d$, $\Gamma$ and $\xi \in \mathbb{R}^{2N}$, the symmetrically ordered characteristic function of the Gaussian state becomes
\begin{equation}
\begin{split}
 \chi(\xi)& =\mbox{Tr}(D(\xi)\rho) \\
          & =\mbox{exp}\left(i d^T\xi-\xi^T\Gamma \xi/4 \right),
 \end{split}
\end{equation}
% \begin{equation}
% \begin{split}
%  \chi(\xi)& =\mbox{Tr}(D(\xi)\rho) \\
%           & =\mbox{exp}\left(i \xi.Jd-\xi.J^T\Gamma J\xi/4 \right)
%  \end{split}
% \end{equation}
where $D(\xi)=\mbox{exp}[ir^T\xi]$ is the Weyl displacement operator. 
% Following the Heisenberg uncertainty relation, the covariance matrix satisfy the operator inequality $\Gamma\geqslant iJ$. 
% Now let us turn to quantum operations on Gaussian states. 
An operation on a Gaussian state is called ``Gaussian'' if the evolved state still be a Gaussian state. The Gaussian unitary transformations, $U=D(l)U_S$, can be mapped to the real symplectic transformations on the first and second moments \cite{Arvind95}. For $\rho^\prime=U\rho U^\dag$, the moments evolve to $d^\prime = Sd+l$ and $\Gamma^\prime=S\Gamma S^T$ where $S\in Sp(2N,\mathbb{R})$ is a symplectic matrix which corresponds to the action of $U_S$ on the Gaussian state $\rho$. 
% In this work we constrain ourself to the unitaries whose exponents are quadratic in the canonical operators. 
% evolution of the moments become: $\dot{d}=$
For a smooth dynamical map, QFI can be expressed in terms of $d$ and $\Gamma$ \cite{Jiang14}, as
\begin{equation}
 \mathcal{F}^2(\rho)=\frac{1}{4}\left( \frac{1}{2}\mbox{Tr}(\Phi \dot{\Gamma})+2 \dot{d}.\Gamma^{-1} \dot{d} \right)
 \label{eq:QFIgauss}
\end{equation}
% where $\dot{\Gamma}=\Gamma \Phi \Gamma^T-J \Phi J^T$. 
with $\Phi=(\Gamma \dot{\Gamma} \Gamma^T-J \dot{\Gamma} J^T)^{-1}$, where the pseudo inverse (Moore-Penrose inverse) is considered. In general derivation of $\Phi$ is somewhat tricky. However, a generic form of $\Phi$ can be provided for the class of CMs satisfying the relation: $(\Gamma J)^2=-\nu^2\mathbb{I}$  \cite{Jiang14}. Again, without loss of generality, the first moment $d$ can be removed by a counter displacement: $\rho \rightarrow e^{-ir.Jd}\rho e^{ir.Jd}$. If the unitary involves no displacement, we can remove the second term in Eq. (\ref{eq:QFIgauss}) involving $\dot{d}$, which is the case for the rest of our discussion. Hence the time evolution of the Gaussian state, driven with the Hamiltonian $H$, is dictated by $\dot{\Gamma}=H\Gamma+\Gamma H^T$.
% $\rho \rightarrow \mbox{exp}[-ir.Jd](\rho)\mbox{exp}[ir.Jd]$. 

% \overline{0}
% \textit{ Local quantum Fisher information (lQFI) and quantum correlation} -- 
Consider $1 \times 1$ Gaussian states $\rho \in \mathcal{H}_A \otimes \mathcal{H}_B $ and  party-A follows a local unitary evolution driven with the Hamiltonian $H_A=T(H_a \oplus 0)T$, where $0$ is the $2 \times 2$ null matrix and $[T]_{ij}=\delta_{j,2i-1}+\delta_{j+4, 2i}$. For the Gaussian unitaries which are quadratic in canonical operators, an arbitrary single mode Hamiltonian can be written as $H_a=\overline{m}.\overline{\sigma}$ and the most informative Hamiltonians are with $|\overline{m}|=1$; $\overline{m}=(\cos \theta, \sin \theta \cos \phi, \sin \theta \sin \phi)^T$. Here $\overline{\sigma}=(\sigma_x,-i\sigma_y,\sigma_z)$ are the real symplectic generators in $Sp(2,\mathbb{R})$, where $\sigma_i$s are the Pauli matrices. 
% Hence, $\dot{\Gamma}=H\Gamma+\Gamma H^T$. 
Let us define the minimum of local quantum Fisher information (lQFI), over local unitaries, as
% The quantum correlation, in terms of QFI, is defined as 
\begin{equation}
 \mathcal{Q}^2_A=\mbox{min}_{H_A} \mathcal{F}^2(\rho,H_A).
\end{equation}
 Since the $\mathcal{Q}^2_A$ is due to local unitary evolution applied on one of the subsystems, the symmetry %between two subsystems 
is broken and in general $\mathcal{Q}^2_A\neq \mathcal{Q}^2_B$ except for symmetric states. It has several interesting properties such as: (i) invariant under local unitary operations i.e.,  $\mathcal{Q}^2_A(U_A\otimes U_B \rho U_A^\dag \otimes U_B^\dag)=\mathcal{Q}^2_A(\rho)$; (ii) it is convex i.e., non-increasing under classical mixing; and (iii) monotonically decreasing under CPTP maps on B-party, inheriting the properties from Bures distance and QFI \cite{Petz96}. These compelling properties are the essential criteria for a good quantum correlation measure.
% of $\mathcal{Q}^2_A$ fulfill the eligibility criteria for a quantum correlation measure.  
Being qualified with these characteristics, in the following we demonstrate that the \textit{$\mathcal{Q}^2_A$ is  a valid quantum correlation measure}.

We start our analysis by considering the two-mode Gaussian state whose covariance matrix (CM) is
\begin{equation}
 \Gamma=T\begin{pmatrix}
          \alpha & \gamma \\
          \gamma^T & \beta
         \end{pmatrix} T
\label{eq:Gamma}         
\end{equation}
where 
% $T_{ij}=\delta_{j,2i-1}+\delta_{j+4, 2i}$ with 
$\alpha=a\mathbb{I}$, $\beta=b\mathbb{I}$ and $\gamma=\mbox{diag}\{c,d \}$. 
% The $T$ matrix is the transformation matrix from $(x_A,p_A,x_B,p_B)\rightarrow (x_A,x_B,p_A,p_B)$. 
The symplectic invariants of CM  are $I_{1}=\mbox{det}\alpha$, $I_2=\mbox{det}\beta$, $I_3=\mbox{det}\gamma$ and $I_4=\mbox{det}\Gamma$. The CM corresponds to a physical state iff $I_1, I_2 \geqslant 1$ and the symplectic eigenvalues $\nu_{\pm} \geqslant 1$, where $\nu^2_{\pm}=\frac{1}{2} (\Delta \pm \sqrt{\Delta - 4 I_4})$ with $\Delta=I_1+I_2+2I_3$. A Gaussian state is entangled iff the smallest symplectic eigenvalue ($\widetilde{\nu}_{-}$) of partially transposed CM is less than one ($\widetilde{\nu}_{-}<1$), which is obtained by replacing $I_3\rightarrow -I_3$ (i.e., by time reversal) \cite{Simon00}. While our discussion on $\mathcal{Q}^2_A$ as the measure of quantum correlation is applicable to  general states, for explicit calculations we emphasize on the relevant subclasses of states for which $(\Gamma J)^2=-\nu^2\mathbb{I}$ with $a=b$ and $c=-d$. For such states the smallest symplectic eigenvalue of partially transposed CM becomes $\widetilde{\nu}_-=|(|a|-|d|)|$. Thus for all $|d|\leqslant |a|-1$ the states are not entangled.
In this case  lQFI becomes $\mathcal{F}^2(\rho)=\frac{1}{8(\nu^2+1)}\mbox{Tr}(\dot{\Gamma}J)^2$ and a closed analytical formula for quantum correlation measure can be given as 
\begin{equation}
 \mathcal{Q}^2_A=\frac{2}{1+a^2-d^2} \mbox{min}\left[d^2, 2a^2-d^2  \right].
\end{equation}
 Clearly the quantum correlation of a Gaussian state vanishes for $d=0$. This fact is also supported by other measures of quantum correlation such as quantum discord \cite{Giorda10}.  Hence, it reliably quantifies the quantum correlation beyond Gaussian entanglement. Analogously, the maximum of lQFI over all local unitaries is just $\mathcal{P}^2_A=2\mbox{max}\left[d^2, 2a^2-d^2\right]/(1+a^2-d^2)$ which will be useful for later discussion.

\begin{figure}
\centering 
%\vspace{8cm}
\includegraphics[width=0.235\textwidth, angle=0]{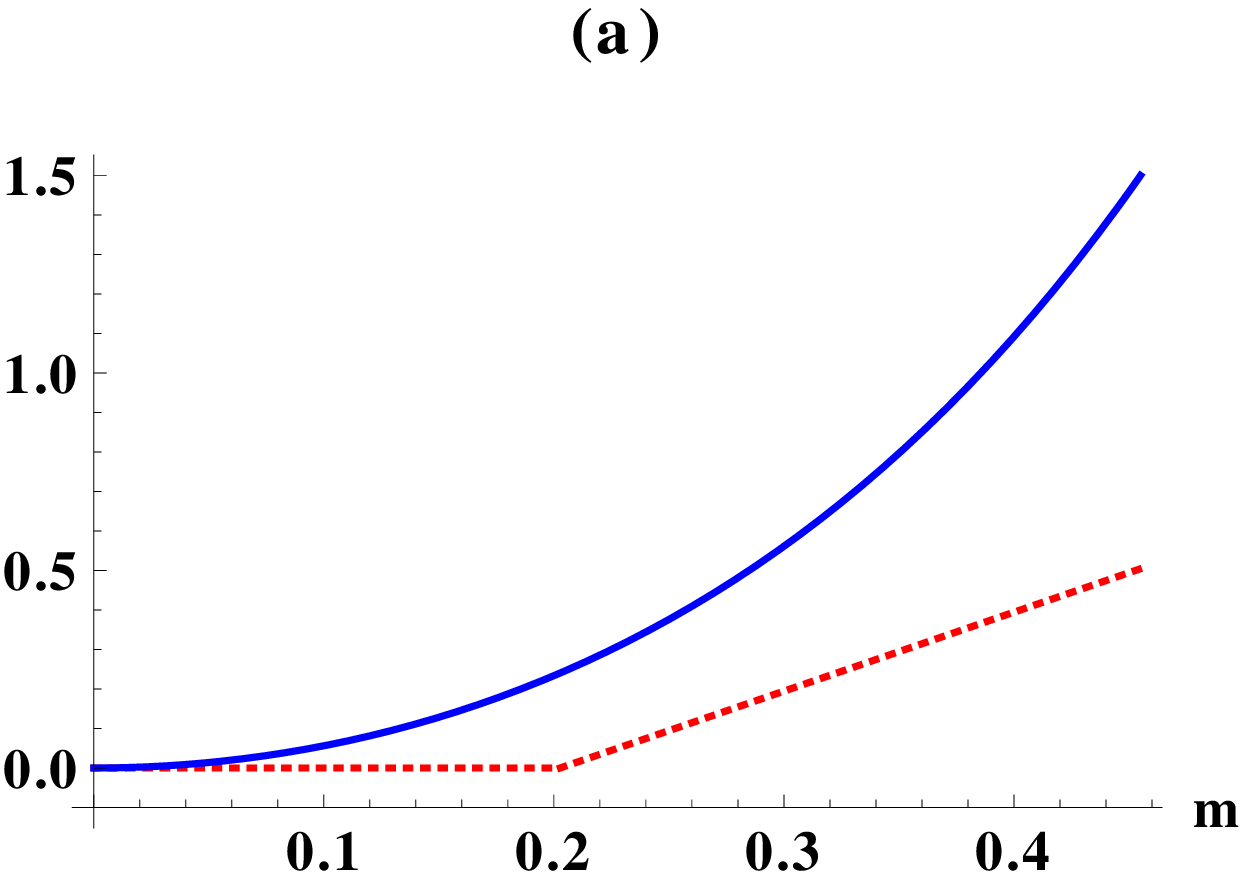}
\includegraphics[width=0.235\textwidth, angle=0]{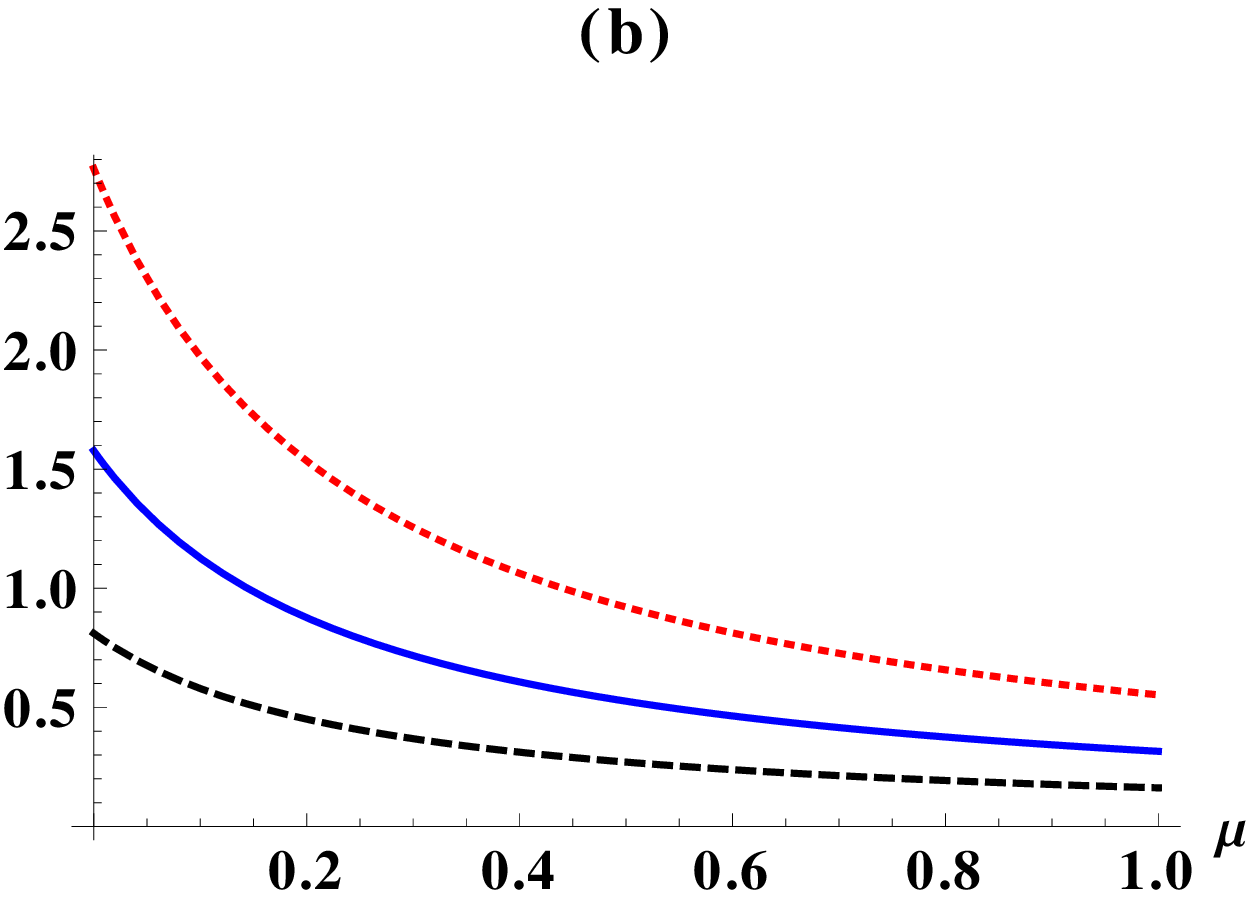}
\caption{\label{fig:EntQcorr} (Color online). The figures depict the characteristics of $\mathcal{Q}^2_A$ for symmetric STS Gaussian states. Fig. (a) represents the variation of quantum correlation $\mathcal{Q}^2_A$ (blue solid trace) and entanglement $\mathcal{N}$ (red dotted trace) with respect to $m$ for a given purity $\mu=1/9$.
% total energy $N_T=4$. 
Fig. (b) shows the variation of $\mathcal{Q}^2_A$ (vertical axis) vs. the purity $\mu$ for fixed $m$. The black (dashed), blue (solid) and red (dotted) traces are correspond to $m=0.3$, $m=0.4$ and $m=0.5$. For details see text.}
\end{figure}

We now illustrate the properties of $\mathcal{Q}^2_A$ using fully symmetric squeezed thermal states (STS) $\rho=D(m)\tau \otimes \tau D(m)$, where $D(r)=e^{m(a^\dag b^\dag -ab)}$ is the two-mode squeezing operator. The $\tau=\sum_k N^k (1+N^k)^{-k-1}|k\rangle \langle k|$ is a chaotic state with $N$ thermal photons in each mode. In terms of $N$ and $m$, the state parameters become $a=b=N_m(1+2N)+N+\frac{1}{2}$ and $c=-d=(1+2N) \sqrt{N_m(N_m+1)}$, where $N_m=\sinh^2 m$. Now we have  $\mathcal{Q}^2_A=\frac{2(1+2N)^2 \sinh^2(2m)}{4+(1+2N)^2}$ and $\widetilde{\nu}_-=\frac{1}{2}(1+2N)e^{-2m}$. To make comparative study between entanglement and quantum correlation we use the logarithmic negativity ($\mathcal{N}$) as the measure of entanglement. For the Gaussian states of our choice it becomes $\mathcal{N}=\max[0,-\ln (\widetilde{\nu}_-)]$ \cite{Adesso05}. In Fig. \ref{fig:EntQcorr} (a) we plot $\mathcal{Q}^2_A$ (measure of quantum correlation) and $\mathcal{N}$ (measure of entanglement) with respect to $m$ for a given purity $\mu=(1+2N)^{-2}=1/9$. The state is separable ($\mathcal{N}=0 $) for $m<0.2$. However,  $\mathcal{Q}^2_A\neq0$ as far as $m\neq0$, i.e., $N_m\neq0$ even in the absence of entanglement and it monotonically increases with $N_m$. So bipartite Gaussian states have nonzero quantum correlation except for product states. Further, the Gaussian states are separable for $N_m\leqslant N^2/(1+2N)$. At the separability threshold $N_m = N^2/(1+2N)$, one easily sees that the $\mathcal{Q}^2_A$ is an increasing function of the total energy ($N_T=4N^2$) of the Gaussian state for fixed $N_m$. Contrarily, the quantum correlation decreases with the increase in purity as shown in Fig. \ref{fig:EntQcorr} (b) which depicts the  variation of $\mathcal{Q}^2_A$ with respect to $\mu$ for a given $m$. For a given purity $\mu=(1+2N)^{-2}$, it turns out that the $\mathcal{Q}^2_A$ increases with entanglement.   
% Fig. \ref{fig:EntQcorr} (b) shows the  variation of $\mathcal{Q}^2_A$ with respect to $\mu$ for a given $m$.
All these properties are comparable with the Gaussian quantum discord and thus justify   $\mathcal{Q}^2_A$ to be the valid measure of Gaussian quantum correlation.

% \textit{ Role in quantum metrology} -- 
Since the proposed measure of quantum correlation is based on quantum Fisher information, an immediate application would be to investigate the role of quantum correlation and, if possible, provide precision bounds in quantum metrology with Gaussian states. 
% For long time researchers believe that, quantum entanglement plays the role to enhance metrological precision \cite{Giovannetti06}. However, recently in \cite{BeraQFICorr}, it has been shown that not only entanglement but also quantum correlation, in discrete variable systems, can be a resource to enhance precision. Now we investigate the same but for Gaussian quantum systems. 
In quantum  metrology, generally, a probe state $\rho$ is driven with unitary evolution such that the evolved state encodes an observable parameter as $\rho_\chi=e^{-i\chi H}\rho e^{i\chi H}$. Here $H$ is the generator Hamiltonian. Then quantum measurements (POVMs) are performed on $\rho_\chi$ to estimate the parameter $\chi$. Interestingly, the lower bound on the error $\varDelta \chi$ in the estimation is independent of the choice of POVMs 
% an unobservable parameter $\theta$ is measured after sending a probe state $\rho$ through unitary evolution such that the evolved state, on which the measurements (POVMs) are performed, is $\rho_\theta=e^{-i\theta H}\rho e^{i\theta H}$. Here $H$ is the generator Hamiltonian. In such process, the $\rho_\theta$ encodes the $\theta$ and it turns out that the lower bound on the error in estimating $\theta$ (i.e., the variance $\varDelta \theta$) is independent of the choice of POVMs.  The lower bound on $\varDelta \theta$ 
and it is solely determined by the quantum Fisher information, and is given by the celebrated quantum Cram{\'e}r-Rao (qCR) bound \cite{Braunstein94},  $\varDelta \chi \geqslant \frac{1}{\mathcal{F}(\rho, H)} $.

Now let us consider quantum metrology with two-mode Gaussian states where one of the modes is driven with local unitary evolution and then measured. In such situation the bounds on metrological error depends on lQFI. Again with non-zero quantum correlation, there exist the tighter bounds on lQFI as $\mathcal{Q}^2_A \leqslant \mathcal{F}^2_A(\rho) \leqslant \mathcal{P}^2_A$  and these eventually lead to tighter qCR bound given by 
\begin{equation}
 \frac{1}{\mathcal{Q}_A} \geqslant \varDelta \chi \geqslant \frac{1}{\mathcal{P}_A}.
\end{equation}
% $ \frac{1}{\mathcal{Q}_A} \leqslant \varDelta \theta \leqslant \frac{1}{\mathcal{P}_A} $. 
Hence the presence of quantum correlation introduces an intrinsic precision that is inversely proportional to the quantum correlation  $\mathcal{Q}_A$ present in the system.

\begin{figure}
\centering 
%\vspace{8cm}
\includegraphics[width=0.3\textwidth, angle=0]{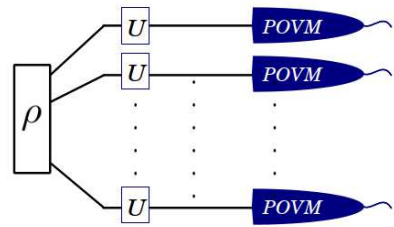}
\caption{\label{fig:Metrology} (Color online). A general scheme for quantum metrology with multi-mode Gaussian quantum state, $\rho$. The $\rho$ is driven with local unitaries ($U=\exp[-iH_k\chi]$) and thus it encodes an unobservable parameter, $\chi$ (a shift in phase). The task is to estimate the $\chi$, with high precision, by performing measurements (POVMs) on the evolved state.}
\end{figure}

The situation becomes very different when quantum metrology involves multi-mode Gaussian states and each mode is driven with local unitary and measured subsequently as shown in Fig. \ref{fig:Metrology}. If the Gaussian states are product states $\rho=\rho^{\otimes n}$, the total QFI (tQFI) is just the sum of lQFIs i.e., $\mathcal{F}^2(\rho,\oplus_1^n H)=n \mathcal{F}^2(\rho,H)$. For product states the metrological precision can be achieved is limited by the standard quantum limit (SQL), $\varDelta \chi \geqslant \frac{1}{\sqrt{n}\mathcal{F}(\rho,H)}$. However, this is not the case when there present quantum entanglement ($\rho \neq \rho^{\otimes n}$), and one is able to go beyond SQL to reach the Heisenberg limit \cite{Paris09}. Here we show that not only entanglement but also quantum correlation can lead one to go beyond SQL.  With the driven local Hamiltonians $H=H_A+H_B$, tQFI becomes
\begin{equation}
\begin{split}
 \mathcal{F}^2(\rho,H)=& \mathcal{F}^2(\rho,H_A) + \mathcal{F}^2(\rho,H_A) \\
                       & + 2 \mathcal{C}(\rho,H_A,H_B),
\end{split}
\end{equation}
where third term is the interference term due to interference between local unitaries with the Hamiltonians $H_A$ and $H_B$. Evidently for  product Gaussian states the $\mathcal{C}(\rho,H_A,H_B)=0$. Consequently, the interference between local unitary evolutions is absent and tQFI is simply the algebraic sum of lQFIs. Thus, for the symmetric bipartite product states and same local Hamiltonians $\mathcal{F}^2(\rho,H)=2\mathcal{F}^2(\rho,H_A)$. Contrarily, in the presence of quantum correlation, $\mathcal{C}(\rho,H_A,H_B)$ can acquire non-zero values and it can implicitly be understood as due to the interference between local unitary evolutions. Without loss of generality, we consider fully symmetric bipartite Gaussian states to explore the properties of the interference term  $\mathcal{C}(\rho,H_A,H_B)$ (although, our discussion can be generalized to arbitrary bipartite Gaussian states). For the fully symmetric $1 \times 1$ Gaussian states 
% For CM, in Eq. (\ref{eq:Gamma}), 
with $a=b$ and $c=-d$ the interference term becomes $\mathcal{C}(\rho,H_A,H_B)=\mbox{Tr}\left[ \dot{\Gamma}_{H_A}J\dot{\Gamma}_{H_B}J  \right]$. We denote $\dot{\Gamma}_{H_A}$ ($\dot{\Gamma}_{H_B}$) as the time derivative of $\Gamma$ solely due to the local Hamiltonian $H_A$ ($H_B$). For the local Hamiltonians $H_A=T(\overline{a}.\overline{\sigma}\oplus 0)T$ and $H_B=T(0 \oplus \overline{b}.\overline{\sigma})T$ where $\overline{a}=(a_x,a_y,a_z)$, $\overline{b}=(b_x,b_y,b_z)$ and $\overline{\sigma}$ are the real symplectic generators in $Sp(2, \mathbb{R})$, the interference term further reduces to $\mathcal{C}(\rho,H_A,H_B)=4(a_xb_x+a_yb_y-c_zc_z)d^2$. Only for the product Gaussian states (i.e., $d=0$) the interference term vanishes identically for arbitrary local Hamiltonians. On the other hand, in the presence of quantum correlation (i.e., $d\neq0$) it can acquire both positive and negative values which is bounded by $ - \mathcal{F}(\rho,H_A) \mathcal{F}
(\rho,H_B) \leqslant \mathcal{C}(\rho,H_A,H_B) \leqslant \mathcal{F}(\rho,H_A) \mathcal{F}(\rho,H_B)$ using the Schwartz in equality. Hence, in the presence of quantum correlation, tQFI is bounded as $(\mathcal{F}(\rho,H_A)-\mathcal{F}(\rho,H_B))^2 \leqslant \mathcal{F}^2(\rho,H) \leqslant (\mathcal{F}(\rho,H_A)+ \mathcal{F}(\rho,H_B))
^2$, even in the absence of entanglement. The equality can be achieved for certain Gaussian states and local Hamiltonians.

\begin{figure}
\centering 
%\vspace{8cm}
\includegraphics[width=0.235\textwidth, angle=0]{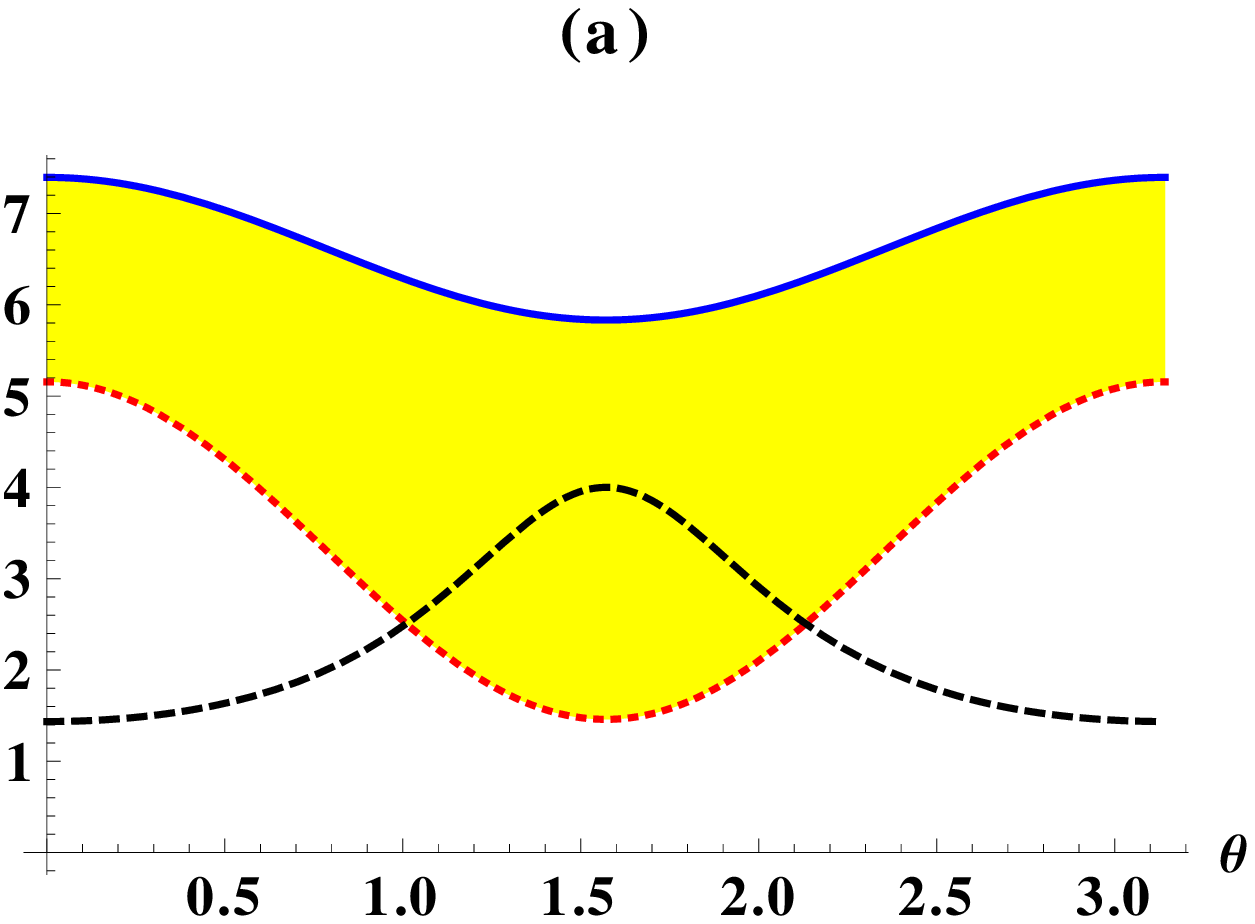}
\includegraphics[width=0.235\textwidth, angle=0]{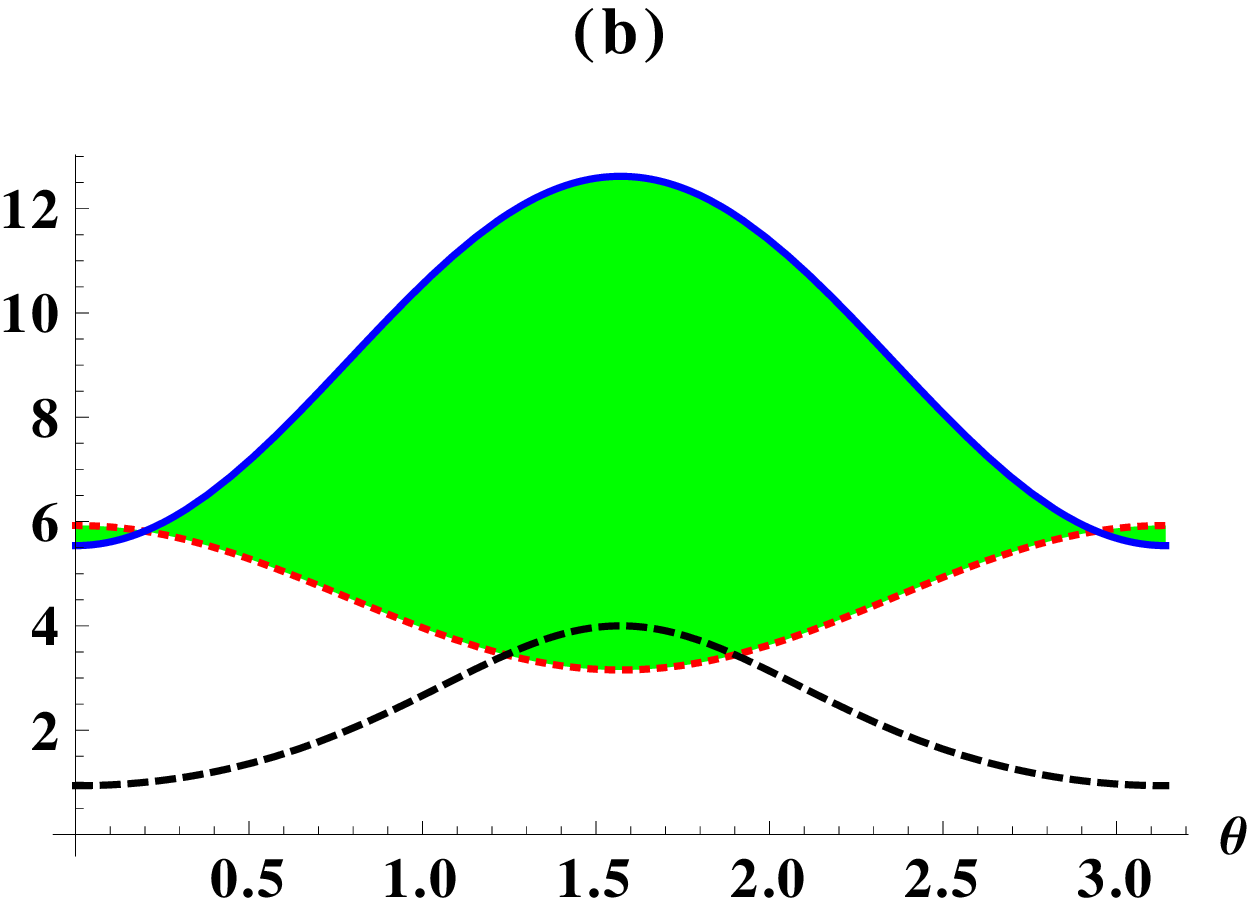}
\caption{\label{fig:QcorrMetro} (Color online). The figures show the variation of lQFI and tQFI for (a) separable but quantum correlated Gaussian states and (b) entangled states vs. $\theta$ for given $\phi=\pi/2$. In both the figures the blue (solid), red (dotted), black (dashed) traces represent tQFI, lQFI and ratio between tQFI and lQFI respectively.}
\end{figure}

For further analysis, we consider two-mode STS Gaussian states, one with finite entanglement and another with separable but having non-zero quantum correlation. The parameters of CMs are chosen to be $a=b=N_m(1+2N)+N+\frac{1}{2}$ and $c=-d=(1+2N) \sqrt{N_m(N_m+1)}$, where $N_m=\sinh^2 m$. The local Hamiltonians considered are 
$H_A=T(\overline{a}.\overline{\sigma}\oplus 0)T$ and $H_B=T(0 \oplus \overline{a}.\overline{\sigma})T$,
% $H_A=\overline{a}.\overline{\sigma}\otimes\mathbb{I}$ and $H_B=\mathbb{I} \otimes \overline{a}.\overline{\sigma}$
where $\overline{a}=(\cos \theta, \sin \theta \cos \phi, \sin \theta \sin \phi)$ and $\overline{\sigma}$ are the real symplectic generators in $Sp(2, \mathbb{R})$. In Fig. \ref{fig:QcorrMetro} (a) we plot the variation of lQFI and tQFI for the state parameters $N=3$, $m=0.4$ and the Hamiltonian parameter $\phi=\pi/2$ with respect to $\theta$. In this case $\widetilde{\nu}_-=1.57$, thus the state is separable but with non-zero quantum correlation $\mathcal{Q}^2_A=1.46$. At $\theta=\pi/2$ we have $\mathcal{F}^2(\rho,H)=4\mathcal{F}^2(\rho,H_A)$.  In Fig. \ref{fig:QcorrMetro} (b) we plot the same but with $N=1$, $m=0.6$ and $\phi=\pi/2$. Hence it is an entangled state with $\widetilde{\nu}_-=0.45$ and $\mathcal{Q}^2_A=3.155$. Again, at $\theta=\pi/2$ we have $\mathcal{F}^2(\rho,H)=4\mathcal{F}^2(\rho,H_A)$. It is interesting to note that we always have $\mathcal{F}^2(\rho,H)=4\mathcal{F}^2(\rho,H_A)$ with certain $H_A$ for which  $\mathcal{F}^2(\rho,H_A)=\mathcal{Q}^2_A$. The above observations clearly 
certify that not only entanglement but also quantum correlation is capable of playing constructive roles to increase tQFI and thus to attain better precision in quantum metrology. Similarly, for n-mode quantum correlated Gaussian states, one can have $\mathcal{F}^2(\rho,H)=n^2\mathcal{F}^2(\rho,H_A)$ for which the metrological precision reaches Heisenberg limit with the corresponding scaling $\varDelta \chi \sim \mathcal{O}\left(\frac{1}{n} \right)$.

% Although the Heisenberg limit can be achieved for n-mode system, but the tQFI $\mathcal{F}^2(\rho,H_A) < n^2\mathcal{P}^2_A$. Rather surprisingly we have $\min[n^2 \mathcal{Q}^2_A, \ n^2 (\mathcal{P}^2_A-\mathcal{Q}^2_A)/2] \leqslant \mathcal{F}^2(\rho,H_A) \leqslant n^2 (\mathcal{P}^2_A+\mathcal{Q}^2_A)/2$.  Consequently, the tighter qCR bounds on metrological precision becomes
% \begin{equation}
% \begin{split}
%  n^2 \min \left[ \mathcal{Q}^2_A, \ \frac{\mathcal{P}^2_A-\mathcal{Q}^2_A}{2} \right] \geqslant \frac{1}{\varDelta \chi^2} \geqslant  n^2\frac{\mathcal{P}^2_A+\mathcal{Q}^2_A}{2}
% \end{split}
% \end{equation}
% However, one should keep in mind that above bounds are restricted to the class of local Hamiltonians which are quadratic in canonical operators and situation is expected to alter for other choices of local Hamiltonians.

% \textit{ Conclusion} -- 
In this work, we have demonstrated that Gaussian quantum correlation can be understood and quantified from quantum dynamical perspective. We have introduced a measure of quantum correlation based on quantum Fisher information which represents the quantum speed of evolution. For bipartite Gaussian states the minimum quantum Fisher information, due to local unitary evolution on one of the parties, reliably quantifies the quantum correlation. Since the proposed measure is based on quantum Fisher information and the metrological error is inversely proportional to such quantity, it provides us the premise to investigate quantum metrology in the presence of quantum correlation. We have shown that in the presence of quantum correlation there can be constructive interference between the local quantum evolutions to increase the total quantum Fisher information and thus better precision in metrology. Even in the absence of entanglement, the quantum correlation has been shown to be an important 
resource to beat standard quantum limit and attain Heisenberg limit.

\end{document}